# Etching Mechanism of Niobium in Coaxial Ar/Cl$_2$ RF Plasma

J. Upadhyay,[1] Do Im,[1] S. Popović,[1] A.-M. Valente-Feliciano,[2] L. Phillips,[2] and L. Vušković[1]

[1]Department of Physics - Center for Accelerator Science, Old Dominion University, Norfolk, VA 23529, USA

[2]Thomas Jefferson National Accelerator Facility, Newport News, VA 23606, USA

The understanding of the Ar/Cl$_2$ plasma etching mechanism is crucial for the desired modification of inner surface of the three dimensional niobium (Nb) superconductive radio frequency cavities. Uniform mass removal in cylindrical shaped structures is a challenging task, because the etch rate varies along the direction of gas flow. The study is performed in the asymmetric coaxial RF discharge with two identical Nb rings acting as a part of the outer electrode. The dependence of etch rate uniformity on pressure, RF power, DC bias, Cl$_2$ concentration, diameter of the inner electrode, temperature of the outer cylinder and position of the samples in the structure is determined. To understand the plasma etching mechanisms, we have studied several factors that have important influence on the etch rate and uniformity, which include the plasma sheath potential, Nb surface temperature, and the gas flow rate.

## I. INTRODUCTION

The accelerating structures made of niobium (Nb) are very important components in superconducting radio frequency (SRF) based particle accelerators. We are developing plasma based etching methods to remove the impurities and mechanically damaged layer from the inner surface of the cavity. This would reduce enormously the economic as well as environmental cost of preparing these cavities. SRF cavities are currently polished by chemical etching methods, which also improve the surface smoothness of the cavity [1]. However, conventional technologies for polishing SRF cavities are expensive and harmful to the environment.

The objective of an etching process in an RF discharge is to remove layers of impurities, oxides and surface contaminants that were accumulated during cavity pre-fabrication. These power-dissipating materials should be removed with the gas flow in the form of volatile compounds. The surface of pure Nb achieved with the etching should have roughness at the level that is comparable to the chemical etching methods. The environmental concerns related to exhaust for chlorine (Cl$_2$) gas are minimal. Argon (Ar) is used as the gas carrier for Cl$_2$, since it is inexpensive [2, 3].

To study plasma etching on the inner surface of a SRF cavity, we have adopted a stepwise approach by choosing smooth coaxial cylinder electrodes as an intermediary geometry, where the outer, grounded electrode is to be etched. The discharge is obviously asymmetric, in the inverse analogy to planar geometry, which is predominant in the fabrication of integrated circuits [4, 5]. While in planar systems the asymmetric discharge arrangement is used for the purpose of increasing ion energy at the processed electrode, the present coaxial configuration is opposite. The electrode to be etched is grounded and the inner electrode is RF power driven.

The next step in developing the cavity etching technology is to use a sample, which is a ring rather than a flat coupon. The ring sample has proven to allow controlled studies of the etching mechanisms in cylindrical geometry of a coaxial asymmetric discharge [2, 3]. In all experiments, one or more ring samples were positioned to fit tightly in the inner surface of the outer electrode, to ensure uniform electric and thermal contact. In this paper, the setup with the ring sample was used as a prototype arrangement that reproduces realistically the etching effects on a cylindrical surface.

To develop the plasma etching method for bulk micro machining of long cylindrical structures made of Nb, an understanding of the etch mechanism of Nb in Ar/Cl$_2$ plasma is required. It is also important to know the nature of the etch mechanism responsible for material removal in order to determine if it is purely physical etching (sputtering), purely chemical etching, or a mixture of both. Another factor in developing this process is the uniformity of etch rate across the cylinder, because the non-uniformity in surface quality along the cavity profile may affect the field distribution.

In this paper, we are presenting a number of experimental results that may help to understand the etch mechanisms in a coaxial Ar/Cl$_2$ RF discharge. Our results show that the etching rate follows the Arrhenius form of dependence on the temperature and it shows some analogy to the etching of Si with Cl$_2$ [4, 5].

Further, by changing the sheath potential at the outer surface of the coaxial discharge using the bias on the inner electrode, the ion bombardment energy on the etched surface is changed and the dependence of the etch rate on the bias potential is measured. This would lead to additional clarification of the etching mechanism.

Etch rate uniformity along the cylinder axis is estimated by comparison of the effect on two Nb rings at different places inside the outer cylinder. The effect of the gas flow rate was also probed by operating with different pumping speed and maintaining the same pressure in the processing chamber.

In the following section we discuss the mechanism of etching by $Cl_2$ in an asymmetric coaxial discharge. The experimental set-up for the present work is described in the third section, and the experimental results are presented and discussed in the fourth section.

## II. ETCHING MECHANISM AND PARAMETERS IN COAXIAL Ar/$Cl_2$ PLASMA

Historically, plasma etching mechanisms have been divided in to four categories [5]: sputtering, chemical etching, ion-enhanced energetic mechanism and the ion-enhanced inhibitor process. Sputtering or physical etching is achieved by bombarding the surface with energetic ions and mechanically removing the material, while chemical etching involves the conversion of the material to a volatile product by chemical reactions between the material and uncharged etchant radicals produced in plasma. Energetic ion assisted etching or reactive ion etching remove material by making the gaseous product in the presence of energetic ions, which would not be possible by uncharged plasma species. In ion enhanced inhibitor etching, inhibitor film induces anisotropy, thus the etching process is spontaneous and purely chemical.

In our coaxial type RF plasma reactor, the inner electrode has a smaller surface area than the outer cylinder whose inner surface has to be etched. Due to the presence of a blocking capacitor in the RF power supply, the smaller surface area electrode will have a self-bias potential. The surfaces exposed to plasma are bombarded by the ions with energy gain in potential difference between bulk plasma and the surface. Time averaged plasma potential $V_p$ and the average ion energy $\epsilon_{ion}$ in a collisionless sheath are related [6] as:

$\epsilon_{ion} = q\,(V_p - V_{bias})$   for inner electrode
$\epsilon_{ion} = q\,(V_p)$                for outer electrode

The plasma potential can be changed by applying a positive DC bias to the inner electrode [6-8], which will change the sheath potential on the outer cylinder. The incident ion energy on the inner surface of the outer cylinder varies due to the change in sheath potential as illustrated in Fig. 1.

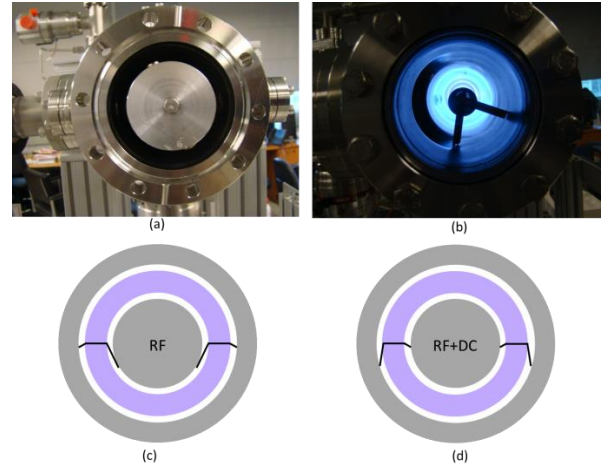

FIG. 1. Cross-sectional images and schematic of coaxial RF electrode, plasma, and sheath potentials: (a) Image of coaxial electrodes (b) Image of coaxial plasma (c) Plasma sheath potential distribution in coaxial RF plasma (d) Plasma sheath potential distribution in coaxial RF plasma with positive DC bias on the inner electrode.

In order to investigate the physical nature of the etching mechanism, we have studied the variation of the etch rate with the sheath potential by raising the DC bias on the inner electrode. To study the chemical nature of the etch mechanism, the temperature of the outer cylinder was varied and the activation energy was evaluated using the Arrhenius equation. The temperature dependence of the etch rate has been studied for a number of materials [9-12] at various plasma conditions. Heating effects on the etch rate for Nb was studied in a $CF_4/O_2$ gas mixture [13] but there was no study for the Ar/$Cl_2$ mixture. The dependence of etching on the flow rate of gases is defined in semiconductor industry by a parameter called residence time. The residence time in a plasma reactor is proportional to pV/F, where p is the pressure, V is the volume of the chamber and F is the gas flow rate, which has an important role in etching characteristics [14]. We investigated the effect of flow rate on the etch rate at the same pressure by variation of the pumping speed employing a partially closed gate valve or by operating only with a mechanical pump.

The dependence of etch rate on the surface area of the material exposed to etching is referred to as a loading effect [15]. In planar type discharges, this effect leads to an etch rate decrease and non-uniformity in etch rate for large area wafer surfaces [15, 16]. This non-uniform etch rate effect in a coaxial type plasma reactor has never been studied. It is much more severe and complex in the case of coaxial discharge because of uni-directional flow of the gases. In planar discharges, used in the semiconductor industry, the gas is flown through a shower head. In that case, all produced radicals are uniformly exposed to the full surface of the material to be etched. In the present design of gas flow in the coaxial discharge, the flow pattern is inhomogeneous, and the materials closer to the flow plume direction get exposed to the fresh radicals first and only the non-consumed radicals can move further. The influence of this effect on Nb etch rate uniformity in coaxial structures has yet to be determined.

## III. EXPERIMENTAL SET-UP

In this study, we have developed a coaxial RF plasma experiment by generating an RF discharge between an inner electrode (varied diameter, 2 to 5 cm) and an outer electrode (diameter a little over 7 cm) with an RF (13.56 MHz) power supply in an Ar / $Cl_2$ gas mixture. The inner electrode was operated with simultaneous RF and DC bias applied. Details of the experimental setup are reported in Ref. [3].

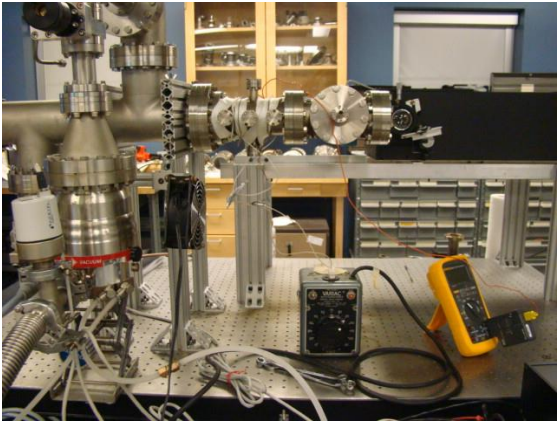

FIG. 2. Photo of experimental set-up for temperature variation during etching cylindrical cavity.

The temperature control of the ring-type sample as shown in Figs. 3(a) and 3(b) is achieved by wrapping heating tapes around the outer wall of the outer cylinder as shown in Fig. 2. The current in the heating tape is controlled through a Variac and temperature is measured by a thermocouple attached to a Fluke multi meter.

The same level of pressure for reduced flow rate is achieved by a gate valve between the vacuum pumping system and the reaction chamber. It is also changed by completely switching off the turbo pump and getting the same pressure by just using the rough pump.

To determine the etch rate variation at different distances from the $Cl_2$ source, we placed two Nb ring-type samples at one inch distance from each other and exposed them to power and gas flow with opposite directions as shown in Fig. 3(c). The ring-type samples are one inch wide and each has surface area of approximately 59 $cm^2$ as shown in Fig 3(a).

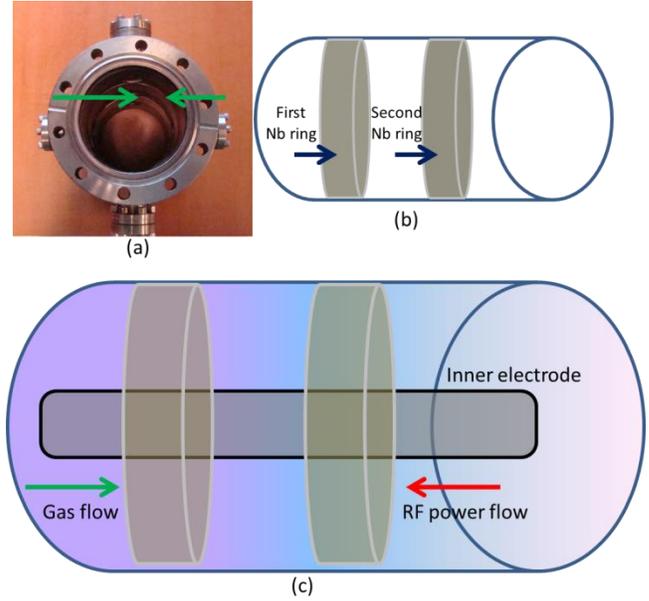

FIG. 3. Schematic and image of the Nb ring sample placed on inner surface of the outer cylinder: (a) Cross sectional view of two ring samples placed on the inner surface of the outer cylinder indicated with arrows (b) Schematic of axial view of the two ring samples (c) Schematic of loading effect with inner electrode and two ring samples when plasma is on. Arrows show the direction of the gas and RF power flow.

Figure 3(c) illustrates the sample configuration and position for the study of the etch rate non-uniformity effect with the presence of inner electrode and plasma. The change in color represents the depletion of radicals due to consumption by the Nb ring positioned close to the gas flow source.

## IV. RESULTS AND DISCUSSIONS

The etch rate data prepared and discussed in this section are obtained by measuring the mass of the ring samples before and after the exposure to plasma, then dividing this mass difference by Nb density and plasma exposed surface area. The calculated error due to measurement of mass and area led to an error in etch rate of 0.82 nm/min, but uncertainties in other process parameters such as in $Cl_2$ concentration (2%), in power (3W), and in pressure (4mTorr), in temperature (2K) and in dc bias (2V) leads to an estimation of about 10% error in the etch rate measurement.

Each experimental run was carried out for more than 90 minutes and in some cases for 100 minutes to avoid the fluctuation in etch rate measurements due to lag time in starting the etching process [17]. New samples are cut from the same Nb sheet for each run to rule out any difference in the material.

### A. Etching rate dependence on temperature and evaluation of activation energy

The temperature of the substrate plays an important role in determining the chemical reaction rates, adsorption of the reactant species to the substrate and desorption of the reaction products from the substrate [11]. It also gives us an indication about the nature of the etch mechanism as chemical etching mechanisms normally exhibit the Arrhenius type dependence on the temperature [13]. To determine the temperature dependence on etch rate, temperature is varied from 311 K (38ºC) to 644 K (381ºC) with all other etch conditions shown. Plot between etch rate versus 1/T is shown in Fig. 4.

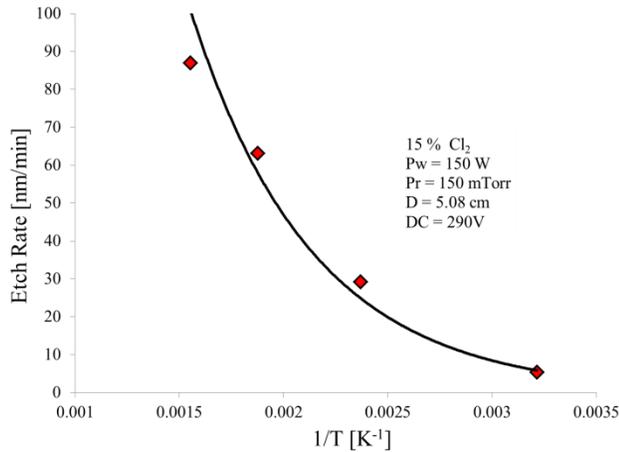

FIG. 4. Etching rate dependence on the Nb temperature.

Figure 4 shows a strong temperature dependence of etch rate, which indicates that the mechanism of Nb etching in Ar/Cl$_2$ plasma has a strong chemical component. Under the condition presented in Fig. 4, the activation energy estimated from the exponential fit of the plot for Nb etch in Ar/Cl$_2$ plasma is 0.15 eV. As higher temperature improves the morphology of the surface [12], particularly at higher temperature with increased power level [9], increasing the temperature of the SRF cavity for faster etching and smoother surface is a viable option.

### B. Etching rate dependence on the DC bias

In plasma processes an ion sheath is formed in front of any material exposed to plasma due to high mobility of the electron. Through this sheath positive ions gain kinetic energy and hit the surface of the material [18]. In our coaxial RF plasma reactor when no DC power supply is attached, the inner electrode acquires a negative self-bias potential (on the order of a couple hundred volts for moderate power of 100 Watt), which varies with power, pressure and gas composition.

A negative bias on the inner electrode does not influence the plasma potential, but a positive bias on the electrode will cause the plasma potential to increase. As all grounded surfaces are bombarded with positive ions with energies characteristic of the plasma potential, by positively biasing the inner electrode, the incident ion energy on the grounded surface can be increased [7]. The variation of the DC bias on the inner electrode and the measured etch rate are shown in Fig. 5.

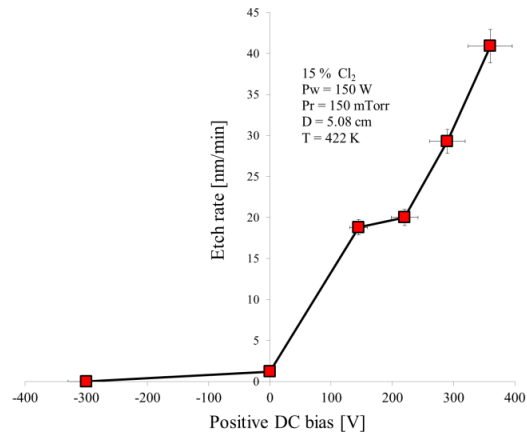

FIG. 5. Etching rate dependence on the DC bias at the inner electrode. Solid line is a visual guideline.

Figure 5 shows a strong dependence of the etching rate on the bias potential (V), and the positive side of the graph shows a very good fit to $V^{0.57}$ which is very close to the square root of the energy fit reported in other works [10, 19]. The knee in the curve might indicates two thresholds for the ion etching, one related to (Ar$^+$, Cl$^+$) and the other threshold for (Cl$_2^+$) as suggested in [19]. This also shows that below a certain bias potential there is no etching possible on the grounded Nb surface, which tells us that a critical amount of plasma potential is required to start the Nb etching process.

### C. Etching rate dependence on the type of gas and other etching parameters

The etch rate dependence on the temperature indicates a strong chemical component, while strong dependence of etch rate on ion energy suggests a physical form of etching. To clarify these results, a new set of measurements was conducted using pure Ar gas to separate the physical component of the etching process. There was no significant mass removal by pure Ar gas. To separate the chemical component of the etching, the experiment was conducted with Ar/ $Cl_2$ mixture with no positive bias on the inner electrode. There was no significant mass removal observed. Therefore, pure chemical etching of Nb samples on the outer electrode is also not possible. There is appreciable etching only in the case, when there is sufficient ion energy available to hit the surface and the mixture of the gas contains some amount of $Cl_2$. Fig. 6 shows this in graphical format. The Nb (placed on the grounded outer wall) etching is possible only in the condition, when we used Ar + $Cl_2$ as a gas mixture and positively bias the inner electrode.

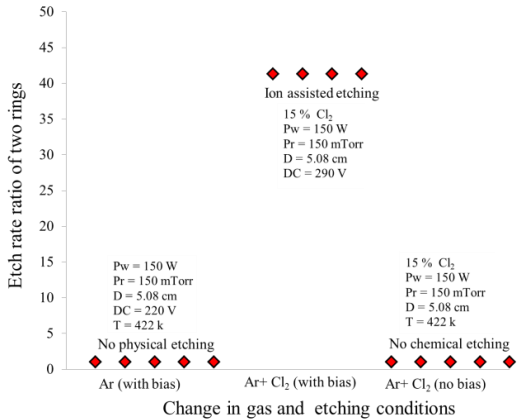

FIG. 6. Etching rate dependence on gas composition and indicated etching parameters.

Figure 6 shows the behaviour very similar to that of Fig. 2 in the paper by Coburn and Winters [20]. It is safe to say that Nb etching in Ar/$Cl_2$ plasma is indeed the reactive ion assisted etching.

The Arrhenius type dependence of the etch rate on the temperature shown in Fig. 4 indicates a strong chemical component of the ion-assisted etching mechanism while strong dependence of etch rate on sheath potential, that is the ion energy shown in Fig. 5 suggests also the presence of the physical aspect of ion assisted etching.

### D. Etching rate dependence on the gas flow rate

Determining the effect of the gas flow rate on the etch rate is a challenging task since the etch rate also depends on the pressure. Two sets of measurements were performed to elucidate this effect. In the first experiment at low pressure range, the flow rate was reduced to one half of maximum and the same pressure was maintained constant with the help of a gate valve to control the pumping speed. The results are given in Table I.

TABLE I. Etch rate variation at fixed pressure and different flow rates achieved using the gate valve.

| Pressure (mTorr) | Flow rate (l/min) | Etch rate (nm/min) |
|---|---|---|
| 60 | 0.25 | 84 |
| 60 | 0.13 | 47 |

Table I shows that increasing the flow rate contributes to increasing the etch rate. The lower etch rate in reduced flow rate condition can be attributed to lower availability of the radicals.

In the second experiment we shut down the turbo molecular pump, and the same pressure was achieved with the roughing pump. Achieving the same pressure with the roughing pump only requires lower gas flow rate. The etch rate data for both pumps with corresponding flow rates are shown in Table II.

TABLE II. Etch rate variation at a given pressure and different flow rates achieved with and without turbo molecular pump.

| Pressure (mTorr) | Flow rate Turbo on (l/min) | Etch rate Turbo on (nm/min) | Flow rate Turbo off (l/min) | Etch rate Turbo off (nm/min) |
|---|---|---|---|---|
| 300 | 0.55 | 29 | 0.10 | 15 |
| 450 | 0.69 | 39 | 0.21 | 31 |

Table II shows that the reduction in flow rate decreases the etch rate even at relatively high pressure, which means that at this pressure and power the etch rate is still determined by the availability of the radicals. Therefore Table I and Table II both show that the etch rate at lower and higher pressure are constrained by the radical production rate.

### E. Etch rate non-uniformity dependence on various process parameters

The etch rate non-uniformity is caused by the depletion of the reactants [16]. In the present experiment, two ring samples were placed at about 2.5 cm apart and the etch rates were measured at both rings with varied plasma conditions. The ring placed the closest to the gas flow source was etched more than the ring placed further downstream. The diagrams shown below describe the ratio of the etching rates on the downstream and upstream rings as a function of the process parameters and are indicative of the effect of an individual etching parameter on the uniformity of the etching rate. For a perfectly uniform etching, the ratio should be equal to 1 (ideal etch rate) and it is also plotted on the graph. In Fig. 7 is shown the variation of this ratio to the temperature of the Nb sample.

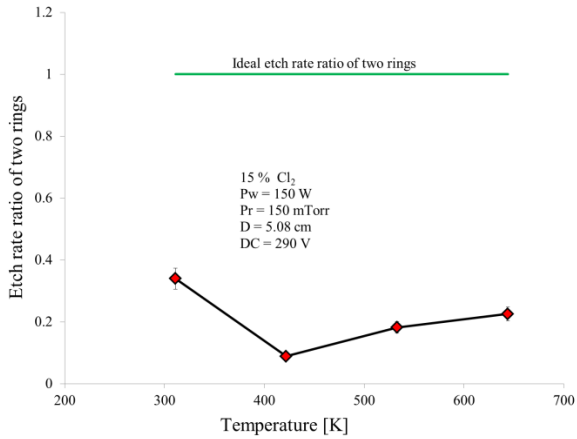

FIG. 7. Etch rate ratio of two rings versus temperature of the substrate. Solid line is a visual guideline.

Figure 7 shows that the increasing temperature does not significantly improve the uniformity of etch rate. It is slightly better at lower temperature. However at lower temperature the overall etch rate is also low as shown in Fig. 4. Therefore we conclude that sample temperature cannot be a very effective control parameter for better uniformity.

In Fig. 8 the variation of the etch rate between Nb rings versus the positive bias on inner electrode is shown. It shows that the etch rate ratio is nominally increasing but at high positively biased inner electrode, when etch rate is usually higher, the ratio decreased.

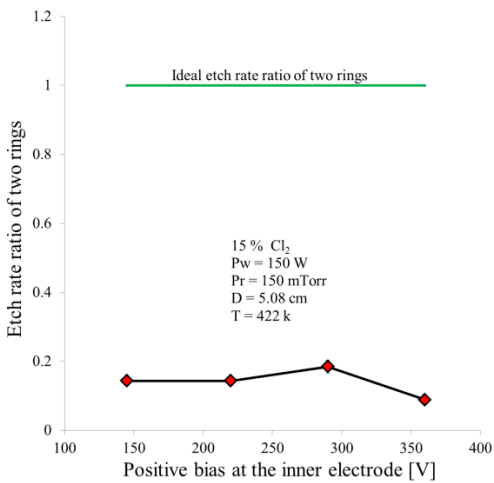

FIG. 8. Etch rate ratio of two rings versus the positive bias on the inner electrode. Solid line is a visual guideline.

In Fig. 9 we present the data for the etch rate ratio and the RF power applied between the electrodes.

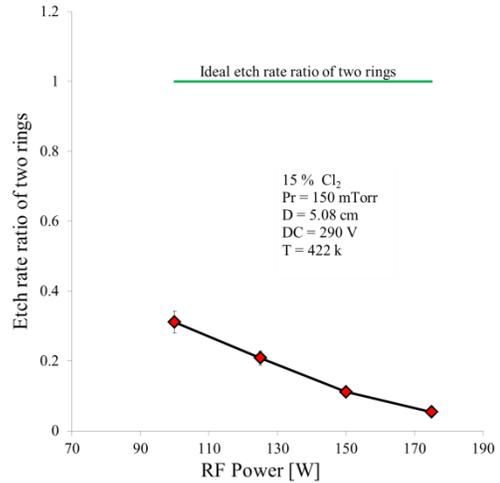

FIG. 9. Etch rate ratio of two rings versus RF power in the system. Solid line is a visual guideline.

Figure 9 clearly shows that increasing RF power reduces the etch rate ratio, which means at higher power when the etch rate is high due to increase of the radical and ion production, the consumption ratio by the first ring also increases. All parameters (temperature, DC bias at inner electrode, RF power) make the reaction probability high for making volatile product. This makes etch rate go high but simultaneously the probability of the radicals to get consumed by the first ring also increases which makes etch rate uniformity decrease, as shown in Figs. 7-9.

Measured etch rate uniformity data for the pressure variation are presented in Fig. 10.

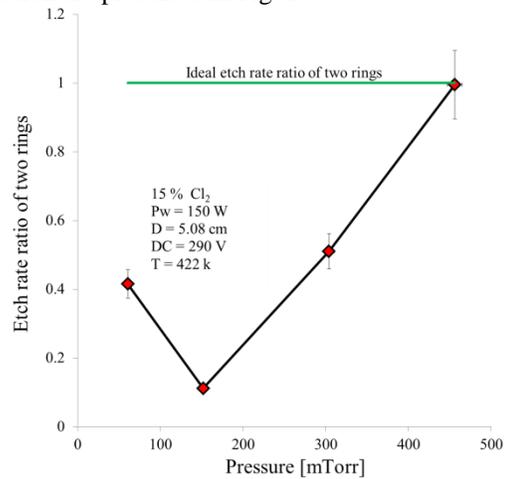

FIG. 10. Etch rate ratio of two rings versus pressure of the system. Solid line is a visual guideline.

Figure 10 shows that when we increase the pressure, which in this case is increased by increasing the flow rate of the gas, the radical density increases. This leads to a better etch rate uniformity. Although the etch rate in the system decreases with the increase in pressure, it does provide the etch rate uniformity very close to ideal.

The data for etch rate uniformity versus the percentage of Chlorine in gas mixture is presented in Fig. 11.

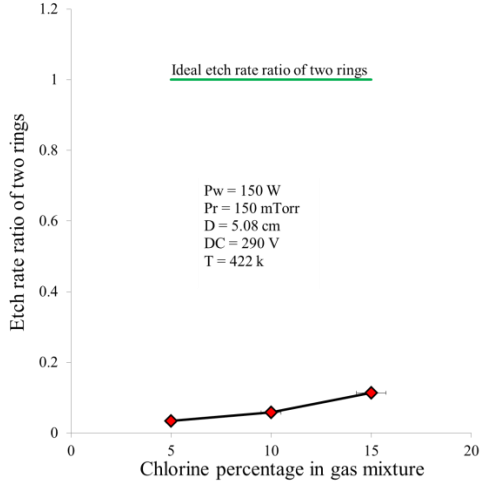

FIG. 11. Etch rate ratio of two rings versus Chlorine percentage in Ar/Cl$_2$ gas mixture. Solid line is a visual guideline.

The data in Fig. 11 show that the etch rate uniformity increases with the increase of chlorine content in the Ar/Cl$_2$ gas mixture. This suggests that etch rate uniformity is constrained by the radical amount in the plasma. We conducted experiment to increase the radical production by changing the diameter of the inner electrode. By decreasing the diameter of the inner electrode we increased the volume of the plasma. The data for etch rate uniformity versus diameters of inner electrode is presented in Fig. 12.

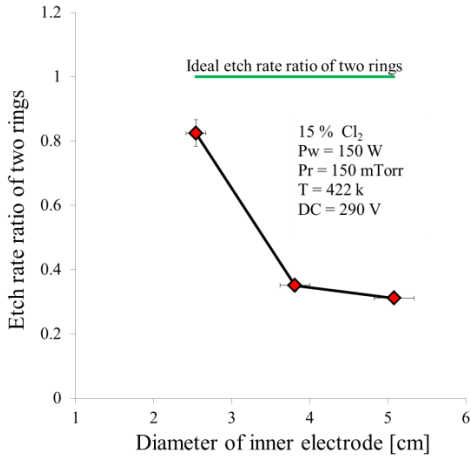

FIG. 12. Etch rate ratio of two rings versus diameter of the inner electrode in the system. Solid line is a visual guideline.

Figure 12 shows that by decreasing the diameter of the inner electrode we can increase the volume of the plasma production region. This increases the chances of the radicals to go further instead of reacting at the first ring only, which makes the etch rate uniformity better. However, uncontrolled reduction of the diameter would lead to low etch rates.

Figures 10-12 show that the parameters which are responsible for the higher radical production (pressure, chlorine concentration, volume production of the radical) also control the etch rate uniformity. The non-uniformity of etch rate in a coaxial plasma reactor is severely constrained by radical consumption by the material closest to the gas flow direction. To verify this effect we removed the first ring and measured the etch rate of the second ring. The results are shown in Table III.

TABLE III. Etch rate variation of the second ring with/without the first Nb ring. Etching conditions are the same.

|  | First ring Etch rate | Second ring Etch rate |
|---|---|---|
| Two Nb rings present | 31 nm/min | 11 nm/min |
| One Nb ring present |  | 37 nm/min |

The table shows that the etch rate of the second Nb ring increases substantially when we remove the first Nb ring. This result confirms that decrease in the second ring etch rate is due to the consumption of radicals by the first Nb ring. The experiment for etch rate non-uniformity effect shows that parameters involved in increasing the number of radicals at the surface, tend to improve the etch rate uniformity.

## V. CONCLUSION

In view of the complex technological challenges that are facing the etching of three dimensional Nb structures, the understanding of the etch mechanism and etching parameters is very important. We offer the following concluding remarks:

i. Nb etch rate can be increased by increasing the temperature of the Nb. The temperature dependence curve follows the Arrhenius type dependence and indicates the chemical nature of the etching mechanism. The activation energy for the etching reaction in Ar /Cl$_2$ gas plasma is 0.148 eV.

ii. The Nb etch rate can be increased by increasing the positive bias on the inner

electrode of the coaxial plasma as it increases the ion incidence energy on the surface. The almost square root dependence on the bias shows the mechanical impact nature of the mechanism, which indicates that there is a physical component to the etching mechanism.

iii. The temperature, bias, and gas variation shows that Nb etching mechanism is neither purely physical etching nor purely chemical. Therefore we identify it as ion assisted etching, or reactive ion etching.

iv. The impact of the gas flow rate on the etch rate is significant.

v. To uniformly etch a cylindrical structure pressure, gas flow rate, and chlorine concentration have to be high, while the diameters of the inner electrode, RF power, temperature, and DC bias have to be low or moderate.

The experiments on the etching mechanism of three-dimensional Nb structures show certain limitations of the coaxial RF plasma reactor. Etch rate non-uniformity experiments show that it will be challenging to uniformly etch an elongated cylindrical structure using a coaxial capacitively coupled RF discharge. A solution for uniform etch rate is presented in a future publication.

## VI. ACKNOWLEDGMENTS

This work is supported by the Office of High Energy Physics, Office of Science, Department of Energy under Grant No. DE-SC0007879. Thomas Jefferson National Accelerator Facility, Accelerator Division supports J. Upadhyay through fellowship under JSA/DOE Contract No. DE-AC05-06OR23177.

## REFERENCES

[1] P. Kneisel, Nucl. Instrum. Methods **557**, 250 (2006).
[2] J. Upadhyay, D. Im, S. Popović, L Vusković, A.-M. Valente-Feliciano and L. Phillips, "Plasma Processing of Large Surfaces with Application to SRF Cavity Modification," Proc SRF 2013, 580 (2013), Paris, France. ISBN 978-3-95450-143-4
[3] J. Upadhyay, D. Im, S. Popović, A.-M. Valente-Feliciano, L. Phillips, and L Vusković, "Plasma Processing of Large Curved Surfaces for SRF Cavity Modification," Submitted to Phys. Rev. ST Accel. Beams, June (2014).
[4] M. A. Lieberman and A. J. Lichtenberg, *Principles of Plasma Discharges and Material Processing,* (John Willey and sons, New York 1994, ISBN 0-471-00577-0)
[5] D. L. Flamm, Pure & Appl. Chem. **62,** 1709 (1990).
[6] H. M. Park, C. Garvin, D. S. Grimard, and J.W. Grizzle, J. Electrochem. Soc. Vol. **145**, 4247 (1998).
[7] J. W. Coburn and E. Kay, J. Appl. Phys. **43**, 4965 (1972).
[8] K. Kohler, J. W. Coburn, D. E. Horne, E. Kay, and J. H. Keller, J. Appl. Phys. **57**, 59 (1985).
[9] S. J. Pearton, A. B. Emerson, U. K. Chakrabarti, E. Lane, K. S. Jones, K. T. Short, Alice E. White, and T. R. Fullowan, J. Appl. Phys. **66**, 3839 (1989).
[10] W. T. Lim, I. K. Baek, J. W. Lee, E. S. Lee, M. H. Jeon et al, Appl. Phys. Lett. **83**, 3105 (2003).
[11] N. M. Muthukrishnan, K. Amberiadis, and A. Elshabini-Riad, J. Electrochem. Soc. **144**, 1780 (1997).
[12] V. M. Donnely, D. L. Flamm, C. W. Tu, and D. E. Ibbotson, J. Electrochem. Soc.: Solid-State Sci. Technol. **129**, 2533 (1982).
[13] M.-M. Chen and Y. H. Lee, J. Electrochem. Soc.: Solid-State Sci. Technol. **131**, 2118 (1984).
[14] A. A. Ayon, R. Braff, C. C. Lin, H. H. Sawin, and M. A. Schmidt, J. Electrochem. Soc. **146**, 339 (1999).
[15] C. J. Mogab, J. Electrochem. Soc.: Solid-State Sci. Technol. **124**, 1262 (1977).
[16] J. Karttunen, J. Kiihamaki, S. Franssila, Proceedings of SPIE , **4174**, 90 (2000).
[17] J. C. Martz, D. W. Hess, and W. E. Anderson, J. Appl. Phys. **67**, 3609 (1990).
[18] H. Kawata, K. Murata, and K. Nagami, J. Electrochem. Soc.: Solid-State Sci. Technol. **132**, 206 (1985).
[19] V. M. Donnelly and A. Kornblit, J. Vac. Sci. Technol. A **31**, 050825 (2013).
[20] J. W. Coburn and Harold F. Winters, J. Appl. Phys. **50**, 3189 (1979).